\documentclass[letterpaper]{article} % DO NOT CHANGE THIS
\usepackage{aaai2026}  % DO NOT CHANGE THIS
\usepackage{times}  % DO NOT CHANGE THIS
\usepackage{helvet}  % DO NOT CHANGE THIS
\usepackage{courier}  % DO NOT CHANGE THIS
\usepackage[hyphens]{url}  % DO NOT CHANGE THIS
\usepackage{graphicx} % DO NOT CHANGE THIS
\usepackage{natbib}  % DO NOT CHANGE THIS AND DO NOT ADD ANY OPTIONS TO IT
\usepackage{caption} % DO NOT CHANGE THIS AND DO NOT ADD ANY OPTIONS TO IT
\usepackage{amsmath,amsfonts}
\usepackage{textcomp}
\usepackage{booktabs}
\usepackage{multirow}

\newtheorem{definition}{Definition}
\usepackage{algorithm}
\usepackage{enumitem}
\usepackage{xcolor}
\usepackage{arydshln}
\usepackage{textcomp}
\usepackage{algorithm}
\usepackage{algpseudocode} % Or \usepackage{algorithmic} if you prefer
\usepackage{amsmath} % For \operatorname
\usepackage{url}

\usepackage{verbatim}
 
\usepackage{tcolorbox}

\usepackage{newfloat}
\usepackage{listings}
\DeclareCaptionStyle{ruled}{labelfont=normalfont,labelsep=colon,strut=off} % DO NOT CHANGE THIS
\floatstyle{ruled}
\newfloat{listing}{tb}{lst}{}
\floatname{listing}{Listing}
\title{Conversational Learning Diagnosis via Reasoning Multi-Turn Interactive Learning}
\author {
    Fangzhou Yao\textsuperscript{\rm 1},
    Sheng Chang\textsuperscript{\rm 1},
    Weibo Gao\textsuperscript{\rm 1},
    Qi Liu\textsuperscript{\rm 1}\thanks{Corresponding Author.} 
}
\affiliations {
    \textsuperscript{\rm 1} State Key Laboratory of Cognitive
Intelligence, University of Science and
Technology of China\\
    \{fangzhouyao, changsheng, weibogao\}@mail.ustc.edu.cn,
qiliuql@ustc.edu.cn
}

\usepackage{bibentry}
\begin{document}

\maketitle


\begin{abstract}

Learning diagnosis is a critical task that monitors students' cognitive state during educational activities, with the goal of enhancing learning outcomes. With advancements in language models (LMs), many AI-driven educational studies have shifted towards conversational learning scenarios, where students engage in multi-turn interactive dialogues with tutors. However, conversational learning diagnosis remains underdeveloped, and most existing techniques acquire students' cognitive state through intuitive instructional prompts on LMs to analyze the dialogue text. This direct prompting approach lacks a solid psychological foundation and fails to ensure the reliability of the generated analytical text. In this study, we introduce ParLD, a preview-analyze-reason framework for conversational learning diagnosis, which leverages multi-agent collaboration to diagnose students' cognitive state over multiple dialogue turns. Specifically, ParLD comprises three main components: (1) Behavior Previewer, which generates a student behavior schema based on previous states and learning content; (2) State Analyzer, which diagnoses the tutor-student dialogue and behavior schema to update the cognitive state; and (3) Performance Reasoner, which predicts the student's future responses and provides verifiable feedback to support ParLD's self-reflection with the Chain Reflector. They operate sequentially and iteratively during each interaction turn to diagnose the student’s cognitive state. We conduct experiments to evaluate both performance prediction and tutoring support, emphasizing the effectiveness of ParLD in providing reliable and insightful learning diagnosis.

\end{abstract}

\section{Introduction}
\begin{figure*}[t]
    \centering
    \includegraphics[width=1\textwidth]{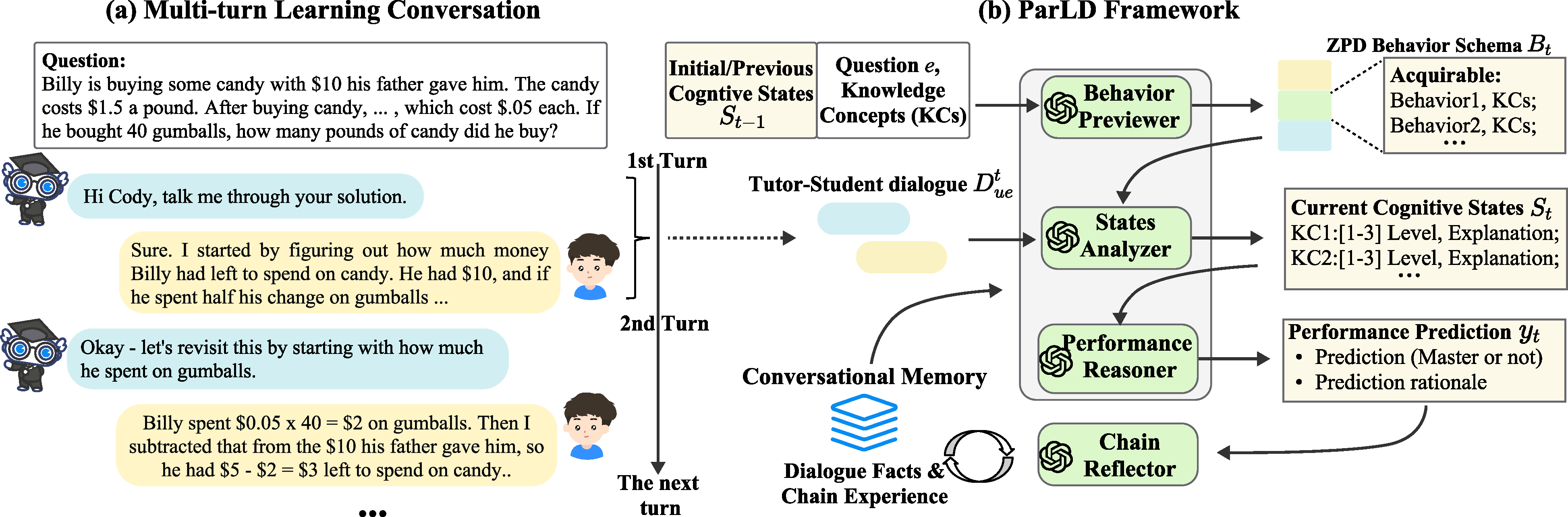}
    \caption{(a) A typical conversational learning scenario where the student interacts with a teacher or intelligent tutor in a turn-based manner.
(b) The ParLD framework for diagnosing cognitive state through the preview-analyze-reason cycle, which can iteratively evolve with each interaction.}
    \label{fig:method}
 
\end{figure*}

% \label{intro}
The rapid growth of educational technology has accelerated the adoption of online learning, prized for its flexibility and personalized experiences. Among various approaches, conversational learning~\cite{long2025conversational,thomas1994conversational,jensen2002conversational} has emerged as a promising paradigm. It enables students to acquire knowledge through interactive dialogue with a human or AI-driven tutor, facilitating tailored guidance and adaptive feedback~\cite{park2024empowering,lv2025genal}. A typical scenario is illustrated in Figure~\ref{fig:method}(a), where teaching unfolds as a multi-turn dialogue centered on solving a given learning objective, often framed as solving a question. At each turn, the tutor adjusts hints, questions, and feedback based on the student’s responses, gradually supporting the learner’s progress by adapting to their current understanding. Effectively supporting this process requires learning diagnosis that continuously monitors the student’s cognitive state~\cite{clow2013overview}, such as their mastery level of key knowledge concepts relevant to the target question. For example, poor performance on an \textit{algebra}-related question signals a need for targeted intervention in that domain.

Despite its promise, accurately assessing students’ evolving cognitive states during multi-turn conversations remains challenging. Traditional methods for modeling students’ cognitive states, such as Knowledge Tracing (KT)~\cite{piech2015deep,ghosh2020context} and Cognitive Diagnosis Models (CDMs)~\cite{lord1980applications,zhang2024understanding}, typically infer knowledge mastery from performance labels like the correctness of students' responses. While effective for discrete exercises, these approaches provide only coarse-grained estimates and fail to capture subtle, continuous cognitive changes occurring within a single problem-solving process. In contrast, learner responses in conversational learning are predominantly open-ended, context-sensitive texts. Cognitive information is distributed and dynamically evolves across multiple turns, making stable signals difficult to extract and limiting the applicability of label-based methods. This calls for fine-grained diagnosis methods to interpret rich textual interactions.
 
Recent advances in Large Language Models (LLMs) offer new opportunities for conversational learning diagnosis. LLMs possess strong language understanding, broad knowledge, and flexible reasoning~\cite{weiemergent,yue2025don}, enabling them to interpret open-ended, context-rich responses and track evolving cognitive states~\cite{laban2025llms,scarlatos2025exploring}. However, existing work primarily applies LLMs to downstream tasks such as adaptive learning or scaffolding, often treating cognitive state analysis as a secondary objective~\cite{liu2024socraticlm,liu2024scaffolding}. These approaches typically perform coarse assessments, analyzing only the final response or the entire dialogue, thereby overlooking the fine-grained, turn-level cognitive dynamics, and potentially introducing bias~\cite{shi2025educationq,echterhoff-etal-2024-cognitive}. Moreover, since cognitive states are latent constructs without observable ground-truth labels, validating LLM-generated outputs remains inherently difficult~\cite{brown2002cognitive,bower2014cognitive}.

To address the aforementioned challenges, we introduce and formulate the Conversational Learning Diagnosis (CLD) task, aimed at analyzing a student's evolving cognitive state within multi-turn tutoring dialogues. To tackle this task, we propose \textbf{ParLD}, an agent-based framework built upon a \textit{\textbf{P}review-\textbf{A}nalyze-\textbf{R}eason} chain. At a high level,  ParLD consists of four core components that work collaboratively: a Behavior Previewer, a State Analyzer, and a Performance Reasoner, which together operate the preview-analyze-reason chain, and a Chain Reflector, which revisits the entire chain to refine the diagnosis results. This framework operates iteratively, with each turn unfolding in a structured sequence. First, inspired by Zone of Proximal Development (ZPD) theory~\cite{shabani2010vygotsky}, the Behavior Previewer projects a behavioral schema based on the student's prior cognitive state and the current learning objective; this schema outlines expected behaviors within their ZPD, providing concrete evidence for the subsequent phase. Next, the State Analyzer scrutinizes the live student-tutor interaction by mapping the dialogue against the projected schema, a process that allows it to update the student's mastery of relevant Knowledge Concepts (KCs). Then, the Performance Reasoner leverages this newly updated cognitive state to predict the student's likely performance on the learning question. Crucially, following each preview-analyze-reason chain, the Chain Reflector initiates a meta-cognitive loop. It reflects on the entire process, cross-references dialogue facts to refine its understanding, and updates an internal experience memory. This self-correction mechanism ensures that ParLD continuously adapts by allowing each component to draw upon the refined memory in subsequent turns, thereby progressively improving the diagnosis's fidelity over time. Our contributions are as follows:
 
 \begin{itemize}[left=1em] 

    \setlength{\labelsep}{0.5em} % 设置标签与内容之间的间隔
    \item We first formulate the Conversational Learning Diagnosis (CLD) task, bridging the critical gap of fine-grained, dynamic tracking of student cognitive state within unstructured, multi-turn dialogues.
    \item We propose \textbf{ParLD}, a novel LLM-based agent framework designed to tackle the CLD task. ParLD's core innovation is its \textbf{\textit{preview-analyze-reason}} chain, augmented by a self-correcting reflective mechanism. This allows the agent to learn from the conversational flow, continuously adapting and improving the fidelity of its diagnosis over time.
    \item We experimentally validate ParLD's effectiveness, demonstrating both high accuracy in performance prediction and the ability to generate impactful tutoring support from its reliable, insightful diagnosis.
\end{itemize}

\section{Problem Definition}
In this section, we begin by defining the core problem and its associated notations to formally address the challenges of conversational learning diagnosis.
% \label{sec:ration}
 
\paragraph{Notations.}
We consider a learning session where a student \( u \) engages in a dialogue with a tutor \( t \) to solve a given question \( e \). This question is associated with a set of knowledge concepts (KCs), denoted as \( K_e \). The entire interaction is captured as a dialogue sequence \( D_{ue} = \{d_1, d_2, \dots, d_T\} \), composed of \( T \) turns. Each turn \( d_t \) consists of a pair of utterances from the tutor and the student. Finally, after the dialogue concludes, the student's overall performance on question \( e \) is recorded as \( r_{ue} \), which can be a binary outcome (e.g., mastered, not mastered) or a qualitative score.

\begin{definition}[\textbf{Conversational Learning Diagnosis}]
\label{def:cla}
Given the dialogue history up to the \( t \)-th turn, \( D_{ue}^t = \{d_1, \dots, d_t\} \), the goal of Conversational Learning Diagnosis (CLD) is to infer the student's latent cognitive state \( S_t \) at that turn. This state, \( S_t \), represents the student's evolving mastery level with respect to each relevant knowledge concept \( k \in K_e \).
\end{definition}

% \begin{definition}[\textbf{Conversational Learning Diagnosis}]  
% Given the multi-turn learning dialogue \( D_{ue} \), the goal of Conversational Learning Diagnosis is to monitor and analyze the student's cognitive state with respect to the relevant knowledge concepts in \( K_e \).
% \end{definition}

% Each dialogue turn \( D^{m}_{ue} \) consists of two components: the tutor's teaching utterance \( d^{m}_{t} \) and the student's learning utterance \( d^{m}_{u} \).
 
The results of the learning diagnosis must accurately reflect the true cognitive state of students, ensuring that they are valuable for both students' self-assessment and the tutor's instructional decisions.
% In the following, we will introduce our proposed framework for the conversational learning diagnosis task in detail.

% \textbf{Conversational learning diagnosis} is concerned with reasoning over conversational data between students and the tutor, with the goal of predicting the performance $\hat{r}_{ui}$ and performing explicit reasoning $Rea_{ui}$.

% This is formally expressed as: $Rea_{ui},\ \hat{r}_{ui} = \text{LLM}(H_u, H_e, d_{ue})$
 
% where $\text{LLM}$ denotes a large language model operating over historical and conversational context.

\section{Methodology}

\label{method}
In this study, we introduce ParLD, a novel multi-agent framework that operationalizes a preview-analyze-reason chain for conversational learning diagnosis. This architecture is specifically designed to infer and model student cognitive state within dynamic, multi-turn conversational learning environments.

\subsection{Overview}
As depicted in Figure~\ref{fig:method}(b), the ParLD framework is architected as a multi-agent system comprising four specialized modules: the \textbf{\textit{Behavior Previewer}}, the \textbf{\textit{State Analyzer}}, the \textit{\textbf{Performance Reasoner}}, and the \textit{\textbf{Chain Reflector}}. These agents operate in an iterative inference loop, processing turn-based dialogue to dynamically model the student's cognitive state. Within each conversational turn, the Behavior Previewer first projects a ZPD-Behavior schema. This projection is conditioned on the prior turn's cognitive state and question text, and the associated KCs. Subsequently, the State Analyzer scrutinizes the live tutor-student dialogue, mapping the interaction against the projected behavior schema to infer and update the student's cognitive mastery of the relevant KCs. Finally, the Performance Reasoner leverages this updated cognitive state to predict the probability of the student successfully answering the learning question. This predictive output is not merely an endpoint. It serves as a critical signal that initiates a reflective process by the Chain Reflector, allowing the ParLD framework to adaptively refine its internal memory and instructional strategy for subsequent interactions.

We provide detailed explanations of each component in the following subsections, and all prompts used in ParLD are available in the code repository.

% predicting examines the underlying rationale behind the student’s final response based on student's assessment results and the final instruction from the tutor; and (3) the Conversation Memory mechanism, caches previewing and analyzing experiences for self-reflecting the generated contents. 

% Note: All the prompts can be referred to in Appendix A.

% Some education assessment research assess student cognitive state with log form data by statistic or data-driven parameter
\subsection{Behavior Previewer}
A primary challenge in diagnosing cognitive state from dialogue is that they are latent constructs, not directly observable. Attempting to map high-dimensional, unstructured text directly to a discrete diagnostic label is an ill-posed problem, often yielding unreliable results due to the significant semantic gap. To address this, our framework introduces the \textbf{\textit{Behavior Previewer}}: previewing a set of plausible and discriminative student behaviors before analyzing the $t$-th turn's dialogue. Inspired by the ZPD theory~\cite{shabani2010vygotsky}, which emphasizes the potential for cognitive growth through guided interaction, we formalize this preview as a ZPD-Behavior schema. This schema acts as a structured prior, constraining the subsequent diagnosis. It categorizes potential behaviors into three zones:
\begin{itemize}
    \item \textbf{Mastered:} Behaviors demonstrated based on prior cognitive state.
    \item \textbf{Acquirable:} Behaviors that can be developed with teacher guidance.
    \item \textbf{Inaccessible:} Behaviors the student cannot perform even with guidance.
\end{itemize}
The zone schema is populated with specific behavioral descriptions and their associated KCs. At turn $t$, the Behavior Previewer agent generates this schema, $B_t$, by conditioning an LLM on the prior cognitive state $S_{t-1}$, the current question's features (question text $e$ and KC set $K_e$), and a task-specific prompt, $\mathcal{P}_{b}$:
\[
B_t = \operatorname{LLM}(S_{t-1}, e, K_e, \mathcal{P}_{b}).
\]
The prompt $\mathcal{P}_{b}$ is to instruct the LLM to generate ZPD-Behavior schema, thereby creating a bounded and interpretable hypothesis space for subsequent diagnosis phases.

\subsection{State Analyzer}
% \label{sec:extra}

The \textit{\textbf{State Analyzer}} serves as the core diagnostic engine within the ParLD framework. While other components are designed to support and refine its diagnosis, this module performs the primary function of inferring the student's mastery level for each relevant KC.

The key to its operation is the ZPD-Behavior schema (\( B_t \)), which provides a structured lens through which to interpret the raw dialogue. Instead of analyzing the dialogue in isolation, the State Analyzer maps the student's observed behaviors in the current turn's interaction, \( d_t \), against the predicted behaviors outlined in \( B_t \). 
For instance, if the student's utterances or problem-solving actions align with behavioral evidence described in the \textbf{Acquirable Zone} of the schema, the system can infer a positive shift in mastery for the associated KCs. This inference process is formally executed by prompting an LLM:
\[
S_t = \text{LLM}(S_{t-1}, B_t, d_t, e, \mathcal{P}_{a}).
\]
Here, \( \mathcal{P}_{a} \) is a prompt engineered to instruct the model to perform this evidence-matching and state-updating task. 
The output, \( S_t \), is a structured representation of the cognitive state. 
As illustrated in Figure~\ref{fig:method}(b), it contains key-value pairs for each KC, detailing not only the mastery level (e.g., \texttt{Good}, \texttt{Fair}, \texttt{Poor}) but also a textual explanation for the diagnosis (e.g., \texttt{\{"KC1": \{"level": "Poor", "explanation": "..."\}\}}). This structured, explainable output is crucial for enabling AI tutors to monitor learning trajectories and make informed instructional decisions.

\subsection{Performance Reasoner}
\label{sec:class}
While the ZPD-Behavior schema provides strong priors for the \textbf{\textit{State Analyzer}}, the framework's reliability needs to be further enhanced by incorporating a verifiable feedback loop. This is the primary function of the \textit{\textbf{Performance Reasoner}}, which enables the Reflector to perform chain reflection on its diagnostic results.

Specifically, the Performance Reasoner takes the State Analyzer's output, the cognitive state \( S_t \), and uses it to predict the student's final performance (\( r_{ue} \)) on the question \( e \). This predictive task is formulated as:
\[
y_t = \operatorname{LLM}(S_{t}, e, \mathcal{P}_{r}).
\]
Here, $\mathcal{P}_{r}$ represents a prompt specifically designed to elicit predictive reasoning from the LLM. 
The resulting output, $y_t$, is a structured tuple formatted as $(\hat{r}_t, \text{Rationale})$, 
where $\hat{r}_t$ is the predicted learning outcome for the current turn 
(e.g., $\in \{\text{mastered}, \text{not mastered}\}$), 
and ``Rationale" provides the textual justification for this prediction.

Crucially, this prediction is verifiable. Once the student's actual performance, \( r_{ue} \), is observed at the end of the session, it can be compared against the prediction \( \hat{r}_t \). 
This comparison provides a concrete error signal that is essential for the framework's self-reflection and refinement, which will be detailed in the next section.

\subsection{Chain Reflector via Memory}
To enable adaptation in a single learning turn, ParLD is equipped with a memory system and a reflective mechanism.
\subsubsection{Conversation Memory.}

The \textbf{Conversation Memory}, \( \mathcal{M} \), serves as an episodic buffer for the current learning session. 
It is designed to store a complete record of the operations within each turn, which we denote as a ``turn trace", \( h_t \).

At its core, this trace contains the dialogue from the current turn \( d_t \), the generated ZPD-Behavior schema \( B_t \), and the inferred cognitive state \( S_t \). 
Crucially, if a reflection is triggered by the Chain Reflector during this turn, the resulting \texttt{R\_trace} is also appended to \( h_t \). After the operations of turn \( t \) are complete, its trace \( h_t \) is added to the memory. 
This update process is formally represented as:
\[
\mathcal{M}_t = \mathcal{M}_{t-1} \cup \{h_t\}.
\]

It is noted that conversational memory is ephemeral and is purged when a new learning conversation begins at a low storage cost.

\subsubsection{Chain Reflector.}

% In that, the Reflector with the Conversation  Memory can ensure that the diagnostic model continuously adapts, improving the fidelity of its cognitive state diagnosis in subsequent turns.\
The \textit{\textbf{Chain Reflector}} is a critical component that drives ParLD's self-correction process. It is activated when a discrepancy occurs between the Performance Reasoner's prediction and the observed student performance at each interaction turn. Upon detection of such a discrepancy, the Reflector systematically revisits the entire preview-analyze-reason chain, querying the Conversation Memory to identify the root cause of the error. 

For example, when auditing the preview-analyze sub-chain at turn \( t \), the Reflector might ask: ``Was the cognitive state \( S_{t} \) correctly inferred, given the dialogue \( d_{t} \) and the schema \( B_{t} \)?" This inquiry is guided by a specific prompt, \( \mathcal{P}_{\text{reflect}} \), which generates a structured critique:
\[
\mathcal{R}_{P \rightarrow A} = \operatorname{LLM}(M_{t}, \mathcal{P}_{\text{reflect}}).
\]
Here, \( \mathcal{P}_{\text{reflect}} \) is the reflection prompt that justifies whether the ZPD-Behavior schema \( B_t \) accurately informs the cognitive state \( S_t \). The \( \mathcal{R}_{P\rightarrow A} \) is a structured output containing two key pieces of information: a \texttt{judgment} (e.g., accurate or not) and a \texttt{critique} (a textual explanation for the judgment). If the reflection indicates that the cognitive state was inaccurately inferred, the Chain Reflector triggers the State Analyzer to rerun the diagnosis process, utilizing \( \mathcal{R}_{P\rightarrow A} \) and the internal memory associated with previous changes. The diagnosis continues iteratively until the performance prediction is accurate. Additionally, we introduce a \textit{max\_num} parameter to control the cost in the experimental setup, limiting the number of reflection iterations.

By leveraging the Chain Reflector via Conversation Memory, this mechanism ensures that the diagnostic model continuously adapts, thereby enhancing the accuracy and fidelity of its cognitive state diagnosis in subsequent interaction turns.
 
\section{Experiments}
\label{expts}
This section details the empirical evaluation of our proposed framework, ParLD. The primary objective of ParLD is to generate accurate, turn-by-turn diagnoses of a student's cognitive state within a conversational learning context. However, evaluating the accuracy of these inferred states is inherently challenging, as the true cognitive state is a latent, unobservable construct. 
Therefore, to validate the efficacy of our framework, we assess its utility through two well-defined proxy tasks that are directly dependent on the quality of the generated diagnoses: (1) student performance prediction and (2) tutoring support. Our experimental design is guided by the following research questions:
\begin{itemize}[leftmargin=*]
\item \textbf{RQ1:} How effective is ParLD at predicting student performance compared to baseline models?
\item \textbf{RQ2:} What are the individual contributions of the Previewer and Reflector modules to ParLD's overall performance?
\item \textbf{RQ3:} What is the utility of ParLD's diagnostic outputs for enhancing pedagogical decision-making in interactive learning scenarios?
\end{itemize}

\subsection{Dataset}

We use the MathDial and CoMTA datasets for experiments:
% MathDial~\cite{macina2023mathdial} consists of 2,861 dialogues between real teachers and students, with the student responses simulated by InstructGPT~\cite{ouyang2022training}, advancing the field of conversational learning. The objective of each conversation is to guide the student toward correctly answering a previously incorrect question. Additionally, each conversation in Mathdial and CoMTA is annotated by teachers to indicate whether the student has fully mastered the question. We adhere to the original train/test split of MathDial and evaluate on the test set. CoMTA~\cite{miller2024llm} is a valuable dataset that records real student interactions with an intelligent tutor from Khan Academy. For our experiments, we filtered 118 conversations with clear learning tasks.
\paragraph{MathDial} \cite{macina2023mathdial} is a large-scale dataset comprising 2,861 dialogues. These dialogues feature interactions between human teachers and a student agent simulated by InstructGPT~\cite{ouyang2022training}. Each conversation is goal-oriented, designed to guide the student toward correctly solving a math problem they had previously answered incorrectly. For our experiments, we strictly adhere to the official train/test split provided with the dataset.
\paragraph{CoMTA} \cite{miller2024llm} provides authentic interaction data, capturing real student dialogues with an intelligent tutoring system from Khan Academy. To ensure a focused evaluation on clear learning objectives, we curated a subset of this data by filtering 116 of 188 conversations that contained clear conversational goals, making them suitable for our turn-by-turn diagnostic analysis.

The conversations in both datasets are annotated with a final label indicating whether the student has fully mastered the question. This valuable label is used to validate the student's performance on the task.
 
\subsection{Utilized LLMs}

Considering the power, need for structured output, and cost-effectiveness of the model, we select \textbf{GPT-4.1}~\cite{openai2025gpt41} and \textbf{GPT-4o}~\cite{openai2024gpt4o} via OpenAI’s API service to construct the ParLD agent framework. All LLMs used in the experiments have their temperature set to \textbf{0} to ensure stable output. Additionally, the maximum reflection time in the Reflector is set to 2 and 1 in MathDial and CoMTa, respectively, to save costs. The code is available at \url{https://github.com/fannazya/ParLD}.
\subsection{Performance Prediction Comparison (RQ1)} 

\paragraph{Motivation.} In this section, we aim to evaluate whether the cognitive state generated by ParLD accurately reflects the students' learning progress. For precise evaluation, we use the annotated labels as the final learning outcome for each conversation, comparing them with the predicted outcomes based on the final cognitive state. Notably, this prediction is made on the final turn without triggering the chain reflection process.

\paragraph{Baselines and Evaluation Metrics.} Given the limited work on the CLD task, we utilize several Knowledge Tracing (KT) models~\cite{scarlatos2025exploring} for comparisons: DKT~\cite{piech2015deep}, AKT~\cite{ghosh2020context}, DKVMN~\cite{zhang2017dynamic}, SAINT~\cite{choi2020towards}, and SimpleKT~\cite{liusimplekt}, which predict students' future performance based on their dialogue learning history. To ensure a fair comparison, we only report the prediction at the final turn of the conversation. The effectiveness of the models is assessed using both classification and regression metrics, including accuracy (ACC) and F1-score.

\begin{table}[t]

\centering
% \setlength{\tabcolsep}{2pt} % Adjust the column separation
% 使用 booktabs 时，建议不在列定义中使用垂直线 |
\begin{tabular}{lcccc}
\toprule
\textbf{Models} & \multicolumn{2}{c}{\textbf{MathDial}} & \multicolumn{2}{c}{\textbf{CoMTA}} \\
\cmidrule(r){2-3} \cmidrule(l){4-5} % 使用 \cmidrule 替代跨列的 \midrule
                 & \textbf{ACC}$\uparrow$ & \textbf{F1} $\uparrow$& \textbf{ACC}$\uparrow$ & \textbf{F1} $\uparrow$\\
\midrule

 DKT     & 58.72        & \underline{65.26}       & 51.88   & 46.40     \\
AKT    & 57.67        & 64.50       & 53.84  &52.88         \\
DKVMN  & 55.15        & 63.04       & 52.31    &       47.06       \\
SAINT   & 58.25        & 63.98       & 51.76    & 47.06  \\
SimpleKT  & 56.70        & 63.38      & 44.58 &  44.73        \\
\midrule % 使用一个 \midrule 来分隔不同的数据组
ParLD (GPT4o) & \underline{65.08}    & 64.04      & \underline{57.02}     & \underline{56.84}     \\
ParLD (GPT4.1) &  \textbf{68.72}     & \textbf{66.15}         & \textbf{57.26}      & \textbf{56.91}     \\
\bottomrule
\end{tabular}
\caption{Comparison of model performance on the MathDial and CoMTA datasets, with the best and second-best models highlighted in \textbf{bold} and \underline{underlined}, respectively. The arrow $\uparrow$ means the higher score, the better
performance. These markers are also for the following results.}
\label{pre}
\end{table}
Table~\ref{pre} demonstrates that ParLD (GPT-4.1) achieves state-of-the-art performance on both the Mathdial and CoMTA datasets, outperforming all traditional KT models. GPT-4.0 also performs well, ranking second in most metrics, except for the F1 score on the Mathdial dataset. Specifically, ParLD (GPT-4.1) surpasses the best-performing DKT model on the Mathdial dataset by a margin of 10\%. This result underscores the potential of LLMs to significantly improve learning diagnosis tasks, particularly in predicting students' cognitive states and the reasoning behind their final predictions. Notably, ParLD (GPT-4.1) outperforms ParLD (GPT-4.0), highlighting how enhanced LLM capabilities can lead to more accurate and insightful predictions of learning progress. This points to a promising direction for future educational technologies.

\subsection{Ablation Study (RQ2)}
To isolate the contributions of our proposed modules, we conducted an ablation study comparing the full ParLD framework against three variants in the performance prediction task. The first variant, \textbf{w/o P}, removes the Behavior \textbf{P}reviewer, thereby analyzing the dialogue directly without the constraining ZPD-Behavior schema. The second variant, \textbf{w/o R}, disables the \textbf{R}eflector, which means memory updates from performance feedback are not incorporated, allowing us to measure the impact of the reflection mechanism. Finally, the \textbf{w/o P+R} variant serves as a baseline without both the Behavior Previewer and Performance Reflector, relying solely on the State Analyzer for performance prediction.

\begin{figure}[t]
    \centering
    \includegraphics[width=0.35\textwidth]{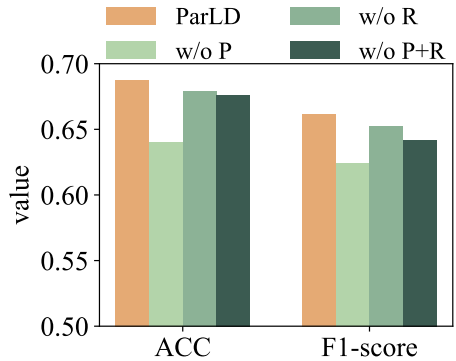}
    \caption{The ablation study of ParLD (GPT-4.1) on the MathDial dataset.}
    \label{abla}
   
\end{figure}

The experimental results in Figure~\ref{abla} demonstrate that, for both GPT-4.1 and GPT-4.0, ParLD outperforms all variants, confirming the contribution of each component to the reliability of its learning diagnosis. Notably, the ``w/o P" variant, which relies solely on the State Analyzer, performs the worst in predicting whether a student has effectively learned the material. This highlights the critical role of the preview-analyze chain, supported by the psychological ZPD-Behavior schema, in providing the necessary context for the State Analyzer. In contrast, reflecting only on the analyze-reason chain can mislead the Analyzer, resulting in less accurate predictions.

\subsection{Effects on Tutoring Support (RQ3)}
In this section, we investigate whether the learning diagnosis results can assist the tutor system in enhancing instructional strategies to improve conversational learning outcomes. Evaluating tutoring support is inherently challenging, as tutoring interventions alter real students’ learning states, making controlled comparisons impossible. To this end, we conduct simulation experiments and perform quality evaluations.

\begin{table}[t]
\centering

\label{inter_transposed}
 
\begin{tabular}{llccc}
\toprule
\multicolumn{2}{c}{\textbf{Model Setting}} & \textbf{CR}$\uparrow$ & \textbf{Avg. T}$\downarrow$ & \textbf{Int. Avg. T}$\downarrow$ \\
% \cmidrule(r){1-2} % 在第1-2列下方划线
\midrule
\multirow{3}{*}{ GPT-4.1} &  -ParLD & \textbf{72.22} &  3.29 & \textbf{2.83} \\
                                 &  -DA    & 62.96 & 3.28 & 3.04\\
                                 &  -DR   & 56.48 & \textbf{3.25} & 3.39 \\
\midrule
\multirow{3}{*}{ GPT-4o}  &  -ParLD & \textbf{62.04} & \textbf{3.43} & \textbf{3.20 }\\
                                 &  -DA    & 53.70 & 3.74 & 3.63 \\
                                 & -DR    & 49.07 & 3.81 & 3.63 \\
\bottomrule
\label{infer}
\end{tabular}
\caption{Comparison of GPT-4.1 and GPT-4o in tutoring support across different settings on the MathDial dataset. The arrow $\uparrow$($\downarrow$) means the higher (lower) score, the better (worse)
performance. }
\end{table}

\subsection{{Tutoring Enhancement Simulation}} To answer \textbf{RQ3}, we designed a simulation experiment to evaluate the pedagogical utility of ParLD's diagnostic outputs. The core motivation is that a more accurate cognitive diagnosis should lead to more effective tutoring instructions, which in turn should help students solve the problem more efficiently. We measure effectiveness by calculating the accuracy of simulated students’ answers to the learned questions when guided by instructions generated from ParLD’s diagnostic results.

\paragraph{\textbf{Simulation Setup.}}We use the same settings as the original dataset, MathDial, simulating students with InstructGPT using the default profile information, and filtering 108 conversations where the simulated student did not answer correctly until the final turn. The simulation proceeds as follows: starting from the second turn, the teacher's utterance is modified by generating improved tutoring instructions based on the cognitive state produced by ParLD. At each turn, the student’s response is verified for correctness; if it is correct, the simulation terminates. We set the maximum number of turns for each conversation to the original number of turns in the dataset. The model settings compared are Direct Respond (\textbf{DR}), which directly responds to the student's response, and Direct Analyze (\textbf{DA}), which simply analyzes the student's cognitive state in each turn and generates corresponding teaching instructions. 

\paragraph{\textbf{Evaluation Metrics.}} We use three metrics as follows: 
\begin{itemize}
    \item \textbf{Correct Rate (CR):} The percentage of simulated dialogues that end with a correct student answer.
    \item \textbf{Avg. Turns (Avg. T):} The average number of turns taken to reach a solution for those dialogues that were successfully solved.
    \item \textbf{Intersection Avg. Turns (Int. Avg. T):} The average number of turns for only those dialogues that were successfully solved by all compared models. This metric provides a fairer comparison of efficiency on a common set of problems.
\end{itemize}
\begin{figure*}[t]
    \centering
    \includegraphics[width=0.9\textwidth]{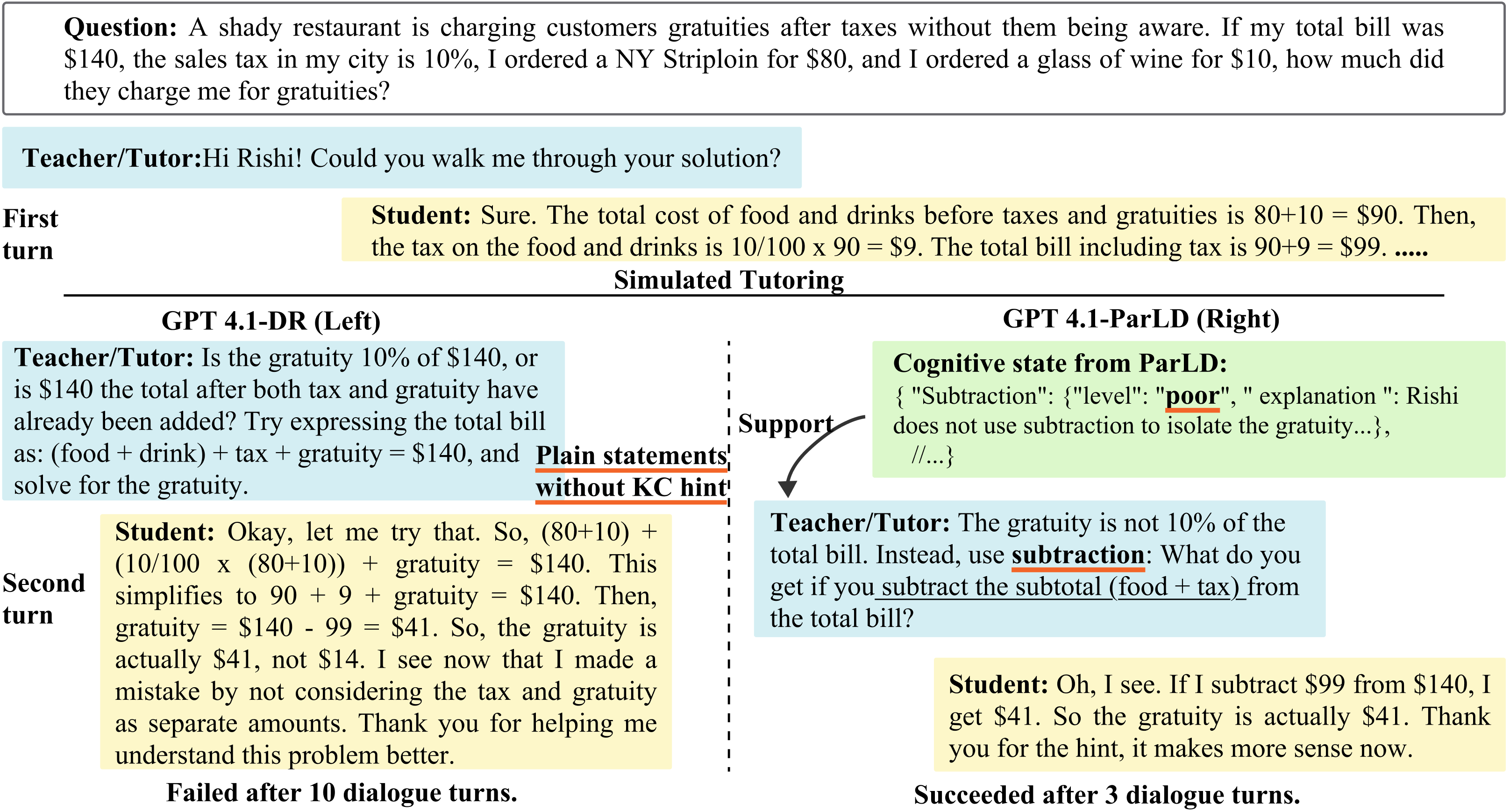}
    \caption{A specific case of multi-turn conversational learning diagnosis and tutoring from the MathDial dataset.}
    \label{case}
 
\end{figure*}
\paragraph{\textbf{Results Analysis.}} 
The results show that ParLD (GPT-4.1) and ParLD (GPT-4.0) significantly outperform both Direct Respond (DR) and Direct Analyze (DA), with DR performing the worst. Specifically, ParLD with GPT-4.1 helps 72.22\% of students effectively learn the material, compared to just 56.48\% with Direct Respond using GPT-4.1. This difference underscores the superior efficacy of ParLD in facilitating effective learning. It also highlights the reliability of the cognitive states diagnosed by ParLD, which are essential for adapting the tutoring process to each student’s needs.

In terms of average dialogue turns (Avg. T), ParLD with GPT-4.0 reduces the number of interactions required for effective tutoring, while yielding results that are similar to those of DR and DA when used with GPT-4.1. This closeness is reasonable because more complex questions naturally require more turns for effective tutoring. The valid advantage of a lower Int. Avg. T on the common set of problems solved by all models using ParLD further supports this point. The reduced Int. Avg. T indicates that ParLD not only improves learning outcomes but also requires fewer dialogue turns to help students master the material efficiently.

In a word, ParLD offers a dual benefit: it helps students overcome learning challenges more effectively while requiring fewer interactions.

\paragraph{\textbf{Quality Evaluations.}} To further evaluate the diagnostic quality of ParLD, we sampled 20 simulated conversations and assessed the reliability of the inferred cognitive states. Two mathematics-proficient students rated the diagnostic results in terms of Accuracy and Coherence using a 3-point scale (1 = worst, 3 = best). ParLD outperformed DA (2.475 vs. 2.1 in Accuracy; 2.7 vs. 2.425 in Coherence), indicating its advantage in conversational learning diagnosis.

\subsection{Case Study (RQ3)}
In this section, we present a specific case of multi-turn conversational learning diagnosis and tutoring from the MathDial dataset to demonstrate the strong learning diagnosis and tutoring enhancement capabilities of our proposed framework. By comparing DR (Direct Respond) and ParLD-supported tutoring (Figure~\ref{case}(a) and (b)), we can observe that ParLD effectively mines evidence of areas where the student struggles, such as \textit{\textbf{Subtraction}}. In the following turn, the tutoring system generates targeted instructions focused on this \textit{\textbf{Subtraction}} KC, helping students identify and fill knowledge gaps, ultimately enhancing learning outcomes. The DR cannot capture the student’s confusion and merely describes the question without specifying the underlying knowledge concept. As a result, the student tutored by DR fails after 10 dialogue turns, whereas ParLD achieves the correct answer in just three turns.

\section{Related Works}

\subsection{Learning Diagnosis}
Learning Diagnosis aims to model and understand student learning processes by analyzing educational data~\cite{clow2013overview}. Within this field, two prominent lines of research are Knowledge Tracing (KT) and Cognitive Diagnosis Models (CDMs). The KT research line~\cite{liu2025deep,gaoKT2025} focuses on predicting student performance by tracking knowledge evolution over time. This area has progressed from foundational models like Bayesian Knowledge Tracing (BKT)~\cite{yudelson2013individualized} to modern deep learning approaches DKT~\cite{piech2015deep} and its advanced variants~\cite{ghosh2020context,shen2021learning,liu2019ekt}. CDMs, conversely, aim to infer students' fine-grained proficiency on specific skills based on their question responses. This includes classic models like Item Response Theory (IRT)~\cite{lord1980applications} and more recent methods like NeuralCDM~\cite{wang2020neural}, which leverages neural networks to model complex student-question interactions. Additionally, recent deep learning-based CDMs have been developed to address practical challenges in online education, such as item scarcity~\cite{chen2023disentangling,yao2023exploiting}, zero-shot diagnosis~\cite{gao2024zero}, and noisy interactions~\cite{yao2024adard}.
A primary limitation of current methods is their fundamental reliance on structured, discrete data, such as right/wrong answers to questions. This makes them inherently ill-suited for the unstructured, open-ended nature of turn-by-turn conversational learning.

\subsection{Conversational Learning}
Conversational learning~\cite{jensen2002conversational,thomas1994conversational} leverages dialogue as a primary mechanism for knowledge acquisition and skill development. Early pioneering works in this area, often categorized as Intelligent Tutoring Systems (ITS), laid the foundational principles. Systems like AutoTutor~\cite{graesser2004autotutor} and other tutorial dialogue systems~\cite{olney2010tutorial} used techniques such as Latent Semantic Analysis (LSA) and scripted dialogue moves to guide students. However, these traditional systems often relied on heavily pre-authored content and rule-based dialogue managers, which made them labor-intensive to build and limited their flexibility.
The advent of Large Language Models (LLMs) has catalyzed a paradigm shift in conversational learning~\cite{milano2023large}, overcoming many of these traditional limitations with their extraordinary knowledge and strong interactivity. Existing research has focused on leveraging LLMs as powerful tutors or educational tools for applications such as Socratic teaching~\cite{liu2024socraticlm,ding2024boosting}, generating dynamic learning scaffolds~\cite{liu2024scaffolding}, and recommending personalized learning paths~\cite{lv2025genal,10.1145/3744367.3744388}, behavior simulation~\cite{zhang2025simulating,gao2025agent4edu}. However, in many of them, learning diagnosis is treated as an intermediate step for a downstream task, rather than a primary research objective. This highlights a remaining gap in leveraging the full potential of LLMs for dynamic, reliable learning diagnosis within conversational settings.

\section{Conclusion}
 
In this paper, we addressed the critical challenge of unreliable and psychologically ungrounded learning diagnosis in conversational learning environments. To tackle this issue, we introduced ParLD, a novel preview-analyze-reason agent framework that leverages multi-agent collaboration for a more robust diagnosis of students' cognitive state. Unlike prevailing approaches that rely on direct and often brittle prompting of language models, ParLD implements a structured, iterative process. By first generating a Behaviour Preview schema, then using a State Analyzer to interpret the live dialogue in context, and finally leveraging a Performance Reasoner for self-reflection, ParLD establishes a more reliable analytical loop. Our experiments, focusing on the practical utility of the diagnosed states, demonstrated ParLD's effectiveness. The cognitive state produced by our framework significantly enhanced the accuracy of student performance prediction and proved valuable for informing appropriate tutoring support. In conclusion, ParLD represents a significant step towards more reliable, insightful, and actionable learning diagnosis, providing a foundational component for the next generation of truly adaptive conversational tutoring systems.

% \begin{table}[h]
% \centering
% \caption{Loss values across different epochs}
% \label{tab:loss_values}
% \begin{tabular}{ c|c|c|c|c|c}
% \hline
% \textbf{Epochs}           & \textbf{2} & \textbf{4} & \textbf{6} & \textbf{8} & \textbf{10} \\ \hline
% \textbf{Rationale Loss}   & 0.717      & 0.238      & 0.055      & 0.027      & 0.017       \\ \hline
% \textbf{Environmental Loss} & 0.865      & 0.934      & 1.282      & 1.553      & 1.626       \\ \hline
% \end{tabular}
% \end{table}
% \section*{Acknowledgements}
% This research was partially supported by grants from the National Key Research and Development Program of China (Grant No. 2024YFC3308200), the National Natural Science Foundation of China (No. 62525606), the Key Technologies R \& D Program of Anhui Province (No. 202423k09020039), and the Fundamental Research Funds for the Central Universities.

\bibliography{aaai2026}

\end{document}